\documentclass{iau}
\usepackage{graphicx}

\title[Modeling line-of-sight magnetograms of active regions] %% give here short title %%
{Modeling line-of-sight magnetograms of emerging active regions}

\author[M. Poisson et al.]   %% give here short author list %%
{M. Poisson$^1$,
  M.~López Fuentes$^1$,
  C.H.~Mandrini$^1$,
  F.~Grings$^1$
 \and P.~Démoulin$^2$
 }

\affiliation{$^1$ Instituto de Astronomía y Física del Espacio (UBA-CONICET), Buenos Aires, Argentina. \\ email: {\tt marianopoisson@gmail.com} \\[\affilskip]
$^2$ Observatoire de Paris (LESIA), Meudon, France.}

\pubyear{2008}
\volume{xxx}  %% insert here IAU Symposium No.
\pagerange{119--126}
\jname{Title of your IAU Symposium}
\editors{A.C. Editor, B.D. Editor \& C.E. Editor, eds.}
\begin{document}

%\usepackage[usenames,dvipsnames]{color}             % For color text: \color command

%\newcommand{\p}[1]{{\color{magenta}{#1}}}
%\newcommand{\pc}[1]{{\color{magenta}{\uppercase{#1}}}}
%\newcommand{\pc}[1]{{\color{ForestGreen}{P: #1}}}
%\newcommand{\mc}[1]{{\color{green}{#1}}}
%\newcommand{\pr}[1]{}
%\newcommand{\fg}[1]{{\color{yellow}{#1}}}
%\newcommand{\pr}[1]{{\color{orange}{\sout{#1}}}}
%\newcommand{\pr}[1]{{\color{yellow}{#1}}}
%\renewcommand{\mp}[1]{{\color{blue}{#1}}}
%\newcommand{\mr}[1]{{\color{blue}{\sout{#1}}}}
%\newcommand{\cc}[1]{{\color{red}{#1}}}
%\newcommand{\mcc}[1]{{\color{magenta}{MC: #1}}}
%\newcommand{\ch}[1]{{\color{orange}{#1}}} 
%\newcommand{\chc}[1]{{\color{JungleGreen}{#1}}}

% General definitions 
%\newcommand{\arcsec}{''} \Bz
\newcommand{\degree}{^\circ} 
\newcommand{\kms}{km s$^{-1}$} %{km/s} 
\newcommand{\Rsun}{$R_{\odot}$}  

% General definitions for derivatives
\newcommand{\f}[2]{{\ensuremath{\mathchoice%
        {\dfrac{#1}{#2}}
        {\dfrac{#1}{#2}}
        {\frac{#1}{#2}}
        {\frac{#1}{#2}}
        }}}
\newcommand{\Int}[2]{\ensuremath{\mathchoice%
        {\displaystyle\int_{#1}^{#2}}
        {\displaystyle\int_{#1}^{#2}}
        {\int_{#1}^{#2}}
        {\int_{#1}^{#2}}
        }}
\newcommand{\curl}{ {\bf \nabla } \times}
\newcommand{\rmd}{{\rm d }}
\renewcommand{\div}[1]{ {\bf \nabla}. #1 }
\newcommand{\grad}{  {\bf \nabla} }
\newcommand{\pder}[2]{\f{\partial #1}{\partial #2}}
\newcommand{\der}[2]{\f{\rmd \, #1}{\rmd \, #2}}

% equations, figures & sections
% equations, figures & sections
\newcommand{\BE}{\begin{equation}}
\newcommand{\EE}{\end{equation}}
\newcommand{\BA}{\begin{eqnarray}}
\newcommand{\EA}{\end{eqnarray}}
\newcommand{\Fig}[1]{Figure~\ref{fig:#1}}
\newcommand{\fig}[1]{Fig.~\ref{fig:#1}}
\newcommand{\figsss}[1]{Figs.~\ref{fig_#1}}
\newcommand{\figs}[1]{Figs.~\ref{fig:#1}}
\newcommand{\figss}[2]{Figs.~\ref{fig_#1} - \ref{fig_#2}}
\newcommand{\sect}[1]{Sect.~\ref{s:#1}}
\newcommand{\app}[1]{Appendix~\ref{app_#1}}
\newcommand{\sects}[2]{Sects.~\ref{sect_#1} and~\ref{sect_#2}}
\newcommand{\eq}[1]{Eq.~(\ref{eq_#1})}
\newcommand{\eqs}[2]{Eqs.~(\ref{eq_#1}) and (\ref{eq_#2})}
\newcommand{\eqss}[2]{Eqs.~(\ref{eq_#1}) - (\ref{eq_#2})}
\newcommand{\eqsss}[3]{Eqs.~(\ref{eq_#1}), (\ref{eq_#2}) and (\ref{eq_#3})}

% This paper specific definitions
\newcommand{\Bo}{B_{\rm 0}}
\newcommand{\dnot}{d_{\rm 0}}
\newcommand{\Nt}{N_{\rm t}}
\newcommand{\yc}{y_{\rm c}}
\newcommand{\xc}{x_{\rm c}}
\newcommand{\minn}{{\rm min}}
\newcommand{\maxx}{{\rm max}}
\newcommand{\phic}{\phi_{\rm c}}
\newcommand{\tauc}{\tau_{\rm c}}
\newcommand{\Fmax}{F_{\rm max}}
\newcommand{\Bz}{B_{\rm z}}
\newcommand{\Cit}{C_{\rm it}}
\newcommand{\Fz}{F_{\rm z}}
\newcommand{\Fzax}{F_{\rm z}^{\rm axial}}
\newcommand{\Fzaz}{F_{\rm z}^{\rm azimuthal}}
\newcommand{\nit}{n_{\rm it}}
\newcommand{\phia}{\phi^{\rm M}_{\rm a}}
\newcommand{\phii}{\phi_{\rm i}}
\newcommand{\phiwl}{\phi^{\rm WL}_{\rm U}}
\newcommand{\phiwlm}{\phi^{\rm WLM}_{\rm U}}

% special letters
%\mathscr (\usepackage{mathrsfs}) and \mathcal defined only for upper case: AB...
%\mathfrak (\usepackage{amsfonts}) defined both for lower and upper cases.   
\newcommand{\Bj}[1]{B_{#1,j}}
\newcommand{\cL}{\mathcal{L}}
\newcommand{\cP}{\mathcal{P}}
\newcommand{\vNabla}{\vec{\nabla}_p}
\newcommand{\vp}{\vec{\mathfrak{p}}}
\newcommand{\vpb}{\vec{\mathfrak{p}_b}}
\newcommand{\Mp}{M_{\vp}}
\newcommand{\Mpi}{M_{\vp_i}}
\newcommand{\Mo}{M_o}
\newcommand{\Mr}{M_r}
\newcommand{\np}{n_p}

\newcommand\maps{\ref@jnl{M\&PS}}% Meteoritics and Planetary Science
\newcommand\aas{\ref@jnl{AAS Meeting Abstracts}}% American Astronomical Society Meeting Abstracts
\newcommand\dps{\ref@jnl{AAS/DPS Meeting Abstracts}}% American Astronomical Society/Division for Planetary Sciences Meeting Abstracts
\maketitle

\begin{abstract}

Active regions (ARs) appear in the solar atmosphere as a consequence of the emergence of magnetic flux ropes (FRs). Due to the presence of twist, the photospheric line-of-sight (LOS) magnetograms of emerging ARs show an elongation of the polarities known as magnetic tongues. These tongues can affect the estimation of tilt angles during their emergence phase. In this work, we propose a Bayesian method to model LOS magnetograms of emerging ARs using a half-torus twisted FR model. We apply this model to 21 emerging ARs observed during Solar Cycle 23. We find that the Bayesian method corrects the tilt when compared to other methods, removing the spurious rotation of the polarities produced by the retraction of the tongues during the emergence. We find a variation in Joy's law with the stage of the AR emergence and the method used for its estimation.   

\end{abstract}

\firstsection % if your document starts with a section,
\section{Introduction}

Active regions (ARs) appear in the solar atmosphere as a consequence of the emergence of magnetic flux tubes (\cite[Fan 2009]{Fan09}). The magnetic field of these flux tubes acquires twist during their formation and rise through the convection zone (CZ), producing the so-called magnetic flux ropes (FRs). Current dynamo models indicate that these FRs originate at the base of the CZ in which the toroidal component of the magnetic field is formed and accumulated (\cite[see Howe, 2009, and references therein]{Howe2009}). Simulations show that instabilities at this depth produce the formation of coherent magnetic field structures that emerge due to the pressure unbalance between the plasma inside and outside these FRs (\cite[Nelson et al. 2013]{Nelson2013}). Once they reach the solar atmosphere they are generally observed as concentrated bipolar magnetic field regions. The bipoles emerge in latitudinal bands between $-35^\circ$ to $35^\circ$ and are slightly inclined with respect to the east-west direction (\cite[van Driel-Gesztelyi and Green 2015]{vanDriel15}). The inclination of the bipoles, known as tilt angle, and its latitudinal dependence, known as Joy's law, are fundamental to understanding the decay and dispersal of the surface magnetic flux (\cite[Wang 2017]{Wang17}).

Simulations relate Joy's law to the effect of the Coriolis force acting over diverging flows near the FR apex during their emergence (\cite[{D'Silva} \& {Choudhuri} 1993]{D'Silva93},\cite[Weber et al. 2023]{Weber23}). The observed dispersion of the tilt, away from the average Joy's law expected behavior, results from the turbulent buffeting on emerging FRs during their transit through the CZ. The strength of the effect of the Coriolis force depends on the duration of the emergence of the FR, and therefore, the observed tilt should decrease with the increase of the FR field strength (\cite[Jha et al. 2020]{Jha2020}). This idea suggests the existence of a tilt quenching in the Babcock-Leighton dynamo that limits the source of the global axial dipole moment.

The tilt angle of ARs is generally estimated using line-of-sight (LOS) magnetograms and computing the flux-weighted center of the polarities, known as magnetic barycenters. With this method, the tilt is defined using the inclination of the line that joints the barycenters with respect to the equator.
Several limitations of this method arise depending on the stage of evolution of the AR producing tilt angles and spurious rotations that may not correspond to the intrinsic inclination of the emerging FR. 
In particular, the twist of the FR produces an elongation of the polarities of the AR observed in LOS magnetograms during their emergence. These elongations known as magnetic tongues were first characterized by \cite{Lopez-Fuentes00}. \cite{Poisson20}
quantified the effect of the magnetic tongues on the estimation of the tilt angle proposing a correction with a method known as the Core Field Fit Estimator (CoFFE). 

In \cite{Poisson22} we developed a method to model the observed LOS magnetograms of bipolar ARs based on a Bayesian inference scheme. The method consists of a 3D model of a half-torus FR from which synthetic LOS magnetograms can be constructed by projecting the vertical component of the field over successive horizontal planes located at different heights. This strictly magnetic field model (no plasma interactions included) with 8-free parameters can reproduce most of the global features observed during the emergence of bipolar ARs (e.g. tilt angle and magnetic tongue extension).

Due to their relevance, tilt angle estimations during the emergence of ARs have to be corrected for the tongue effect. In this work, we aim to model the twist and tilt parameters of the FRs corresponding to the emergence of 21 bipolar ARs. In \sect{method} we summarize the data processing basics and introduce the modeling method. \sect{results} compares the estimations of the tilt obtained with the magnetic barycenters and the Bayesian methods. In \sect{discussion} we discuss the implications of our results.

\begin{figure}[b] \label{fig:AR-8171}
% \vspace*{-2.0 cm}
\begin{center}
 \includegraphics[width=0.95\textwidth]{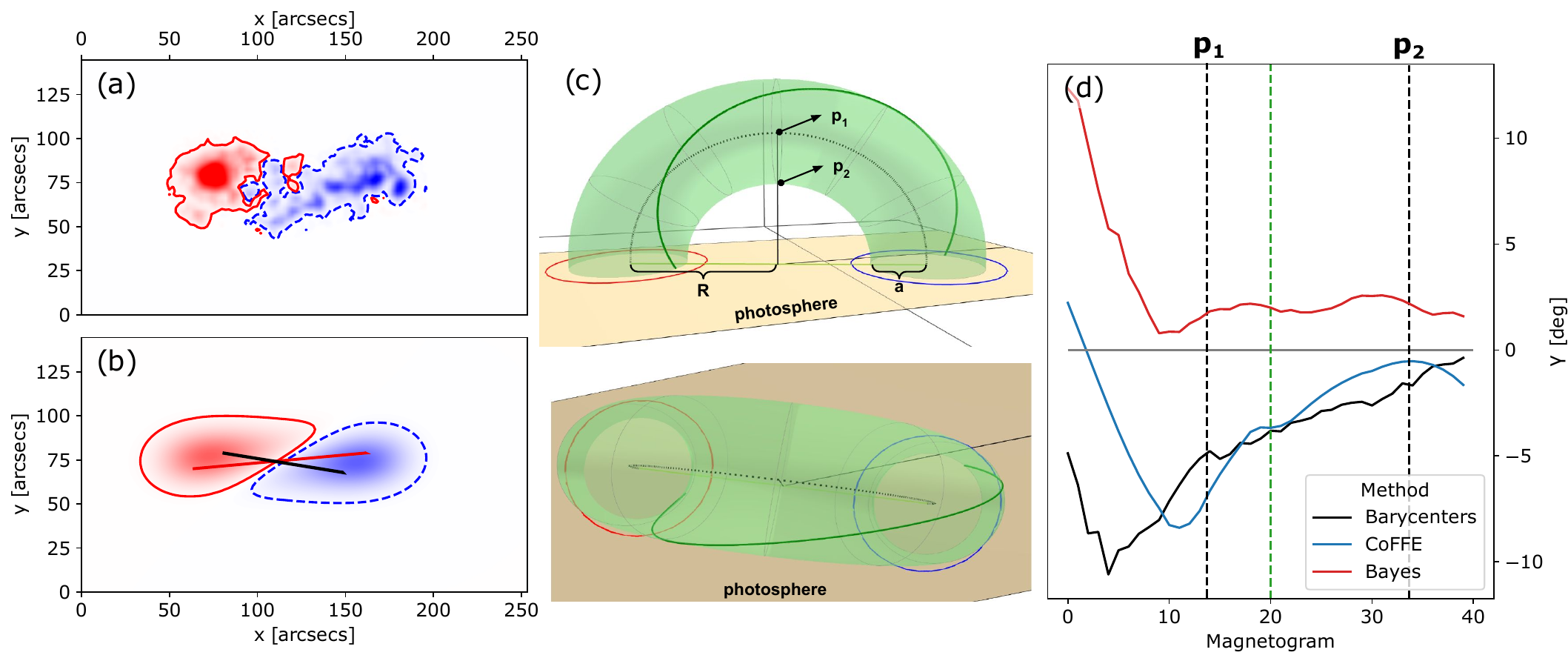} 
% \vspace*{-1.0 cm}
 \caption{(a) SOHO/MDI LOS magnetogram of AR 8171. Red- and blue-shaded areas correspond to the strength of the positive and negative magnetic field. Contours indicate the field strength of $-50$ G (blue dashed line) and $50$ G (red solid line). (b) Synthetic magnetogram of AR 8171 generated with the half-torus model. (c) Top and side views of the half-torus model. In this view, the photosphere height is set to zero. $P_1$ and $P_2$ indicate the relative positions of the photosphere used
for the temporal binning described in \sect{results}. (d) The evolution of the tilt angle of AR 8171 estimated using the magnetic barycenters (black), CoFFE (blue), and the Bayesian (red) methods. The vertical green line corresponds to the magnetogram shown in panel a. Vertical black lines indicate the interpolated times in which the photospheric height is located at $P_1$ and $P_2$ in the Bayesian model. }
\end{center}
\end{figure}

\section{Data processing and modeling method} \label{s:method}

We use LOS magnetograms obtained with the \textit{Michelson Doppler Imager} (\cite[MDI; Scherrer et al. 1995]{Scherrer95}) on board the \textit{SOlar and Heliospheric Observatory} (SOHO). 
The 96-minute cadence magnetograms have a spatial resolution of $1.96''$ and an error per pixel of $9$ G (\cite[Liu et al. 2004]{Liu04}). 

We select 21 emerging bipolar ARs from Solar Cycle 23. We limit the heliographic latitudinal and longitudinal range from $-35^\circ$ to $35^\circ$ with respect to the disk center to avoid projection effects near the solar limb. For each AR, we project the LOS magnetic field into the solar radial direction and rotate the magnetograms to the time the AR is closer to the central meridian using Solar SoftWare tools. We define rectangular boxes that encompass the AR evolution and we mask the background field using a Gaussian smoothing filter and a flux threshold of $20$ G. The parameters of the Gaussian filter are set manually and checked visually for each case to ensure a good selection of the field of the AR. The final output of the processing algorithm provides data cubes containing the magnetograms of each AR. \fig{AR-8171}a shows as an example magnetogram 20 of the evolution of AR 8171. This magnetogram corresponds to the time of the emergence of the AR in which strongly elongated tongues are visible on the leading negative polarity.

We use the same model as the one proposed in \cite{Poisson22}. 
The model is defined by a half-torus magnetic field with four parameters: the small radius $a$, the big radius $R$, the axial flux $\Phi_A$, and the number of turns of the field around the FR axis, defined as $\Nt$. 
Projecting the vertical component of this field into successive horizontal planes, we construct synthetic LOS magnetograms. 
For that, we need at least four other parameters to define the relative position of the FR on each plane. 
These parameters are the horizontal Cartesian coordinates for the torus center defined as $x_c$ and $y_c$, the inclination of the FR axis to the $x-$direction of the plane (the tilt $\gamma$), and the position of the FR apex $d_0$ relative to the photosphere. 
\fig{AR-8171}c shows different views of the half-torus model with the position of the photospheric plane at $d_0 = 1$.

We use a Bayesian inference scheme to compare the synthetic magnetograms with the observed LOS magnetograms of the selected ARs. For that we define the conditional probability distribution (or likelihood function) between the observations and the model as a normal distribution with zero as mean and a standard deviation of $50$ G. The priors are all set as uniform distributions. The ranges for these distributions are defined from direct estimations done over the observed magnetograms. For instance, $a$, $R$, and $\Phi_A$ upper (lower) values are defined as a factor $1.5$ ($0.5$) of the estimated mean size of the polarities, mean separation of the polarities, and maximum magnetic flux, respectively. A detailed description of the application of this method and the definition of priors can be found in \cite{Poisson22}. We use the No-U-Turn sampling algorithm, available in the Python PyMC5 library (\cite[Salvatier et al. 2016]{PyMC3}), to reconstruct the posterior probability distribution. 

We model the full evolution of ARs defining the parameters of the half-torus ($a$, $R$, $\Phi_A$, and $\Nt$) as constant along time. This implies that the full data cube for each AR will have a single possible value for each of these parameters. On the other hand, the position parameters ($x_c$, $y_c$, $\gamma$, and $d_0$) can evolve without any temporal correlation imposed. This method generates a model in which the emergence is driven by a rigid kinetic transport of the FR as a whole across the photosphere. \fig{AR-8171}d shows the evolution of the tilt angle of AR 8171 computed using the magnetic barycenters (black), the CoFFE ( blue), and the Bayesian method (red). In this case, the correction of the effect of the magnetic tongues on the tilt is achieved only with the Bayesian method, the counter-clockwise rotation obtained with the other methods is the effect of the contraction of the tongues.

\begin{figure}[t] \label{fig:tilts}
% \vspace*{-2.0 cm}
\begin{center}
 \includegraphics[width=.95\textwidth]{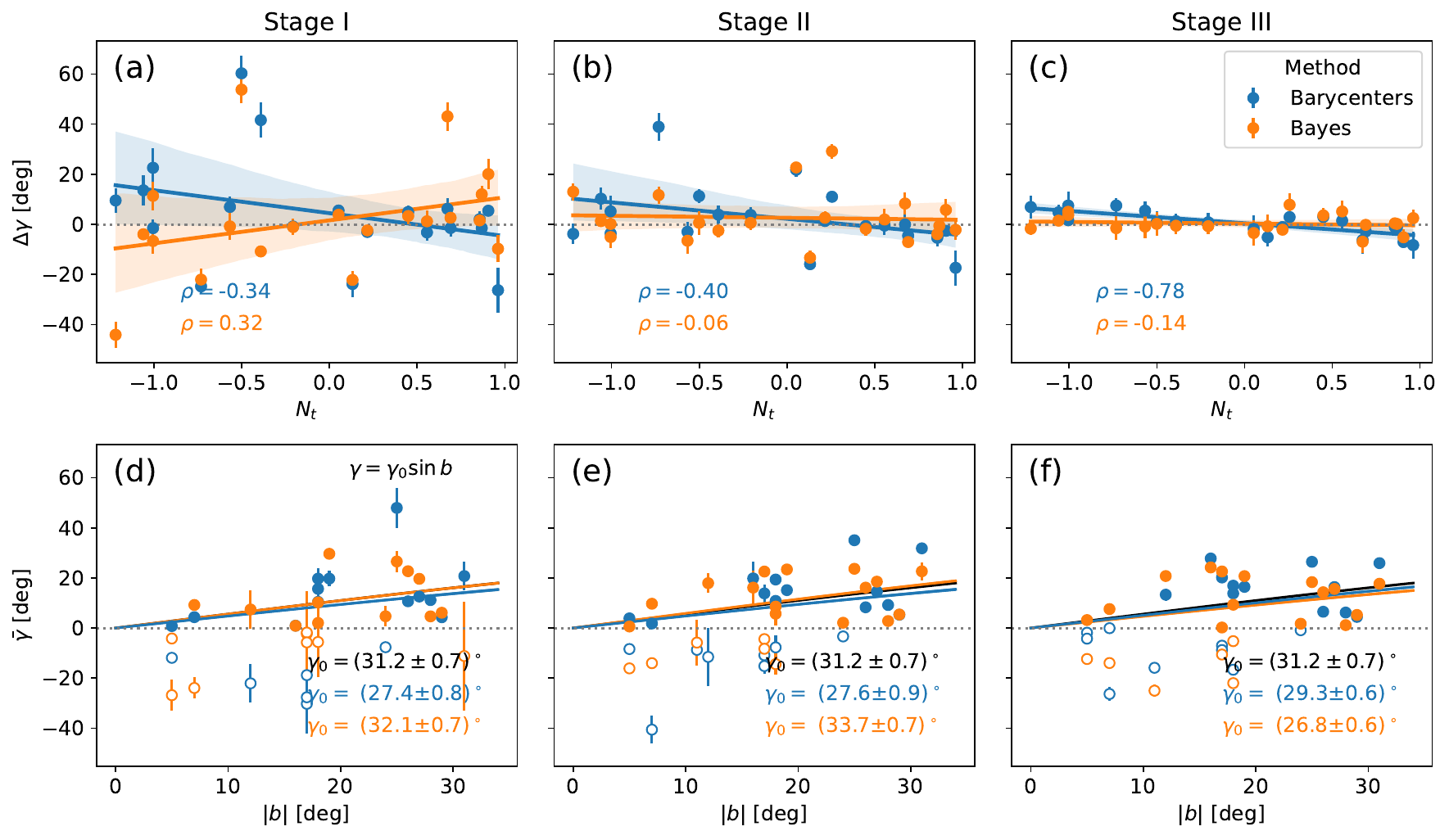} 
% \vspace*{-1.0 cm}
 \caption{(a)-(c) Rotation of the bipoles $\Delta \gamma$ as a function of the number of turns $\Nt$ for the early, mid, and late stages of the evolution of 21 ARs. Blue and orange dots correspond to $\Delta \gamma$ estimated using the magnetic barycenters and the Bayesian method respectively. Linear least-squares fit and correlation coefficients $\rho$ are displayed for each data set. (d)-(f) Mean tilt $\bar{\gamma}$ as a function of the heliographic latitude $b$ for the early, mid, and late stages of the evolution of 21 ARs. We consider both hemispheres with positive latitudes using the absolute value of $b$. The black line corresponds to Joy's law obtained by \cite{Stenflo2012}. Color circles correspond to ARs with negative tilt angles (inclination opposite to Joy's law). Blue and orange lines correspond to the fit of Joy's law for the tilts obtained with the magnetic barycenters and the Bayesian method respectively.}
\end{center}
\end{figure}

\section{Results} \label{s:results}

To compare the tilt at different stages of the AR emergence we separate the evolution of the AR into three time bins. The parameter $d_0$ obtained using the model is a direct estimator of the stage of the FR emergence. Therefore, we divide the emergence using two relative positions of the photosphere: $P_1$ when its depth from the FR apex is equal to the small radius $a$ and $P_2$ when this depth is equal to $2a$ (see marked points in \fig{AR-8171}c). The first bin corresponds to the span between the first observed magnetogram and $P_1$ (stage I). The second bin is defined in the interval between $P_1$ and $P_2$ (stage II). The third bin corresponds to the interval between $P_2$ and the last observed magnetogram (stage III). These bins represent the early, mid, and late stages of the emergence.

Within each of these bins, we compute the variation of the tilt $\Delta \gamma$ and the mean tilt $\bar{\gamma}$. $\Delta \gamma$ has a positive (negative) sign when the bipole rotates clockwise (counter-clockwise). The mean tilt sign is positive when the leading polarity is closer to the equator independently of the sign of the flux or the hemisphere of the AR. \figs{tilts}a-c show the rotation of the bipoles $\Delta \gamma$ as a function of the FR twist $\Nt$ at the three stages of the evolution for 21 ARs. The blue and orange points correspond to the tilt of each AR estimated using the magnetic barycenters and the Bayesian model, respectively. Linear least-squares fits and Pearson's correlation coefficients show the tendencies of the data at each stage of the evolution. 

At stage I the correlation between the bipole rotation and the twist is low for both methods (\fig{tilts}a), meaning that other effects are acting on the bipoles that may produce intrinsic strong rotations apart from the magnetic tongues. Still, differences between both methods at stage I are significant, meaning that both estimations produce different results for the tilt. In the case of the tilt obtained with the magnetic barycenters, stages II and III (\fig{tilts}b-c) show a less dispersed tendency and a direct correlation between the twist and the rotation of the bipoles. This indicates that the contraction of the tongues produces spurious rotations. The rotations are clockwise for the negative twisted FRs and counter-clockwise for a positive twist. This tendency is not found for the tilt obtained with the Bayesian method, since the correlation coefficients of $-0.06$ and $-0.14$, suggest that the tilt angle is corrected from the effect of the magnetic tongues.

\figs{tilts}d-f show the mean tilt as a function of the latitude $b$ at the three stages of evolution for 21 ARs. 
We consider both hemispheres together by taking the absolute value of $b$. The black line marks the Joy's law obtained by \cite{Stenflo2012} in which $\gamma = \gamma_0 \sin b$ with $\gamma_0 = (31.2 \pm 0.7)^\circ$. We fit the latitudinal dependence of the tilt for the analyzed dataset at the different defined stages. We fit only those ARs in agreement with Joy's law, i.e. with positive tilts (solid dots in \figs{tilts}d-f), since the negative tilt cannot be explained by the Babcock-Leighton model and their study requires a different analysis. From this fit, we obtain values of $\gamma_0$ shown in \figs{tilts}d-f. Despite the largest dispersion at stage I,
the tilt angle obtained with the magnetic barycenters produces similar results of Joy's law at all three stages with a mean $\gamma_0$ of approximately $28^\circ$. In contrast, tilt angles obtained with the Bayesian method have values of $\gamma_0$ significantly lower and with lower dispersion at stage III, compared with the earlier stages. This implies that fully emerged ARs relax to a different tilt producing different Joy's law parameters. If we combine all stages as independent observations (similar to what is done when tilt angles are computed automatically from observations without considering the emergence stage of ARs), despite the ARs being repeated, we get $\gamma_0 = (30.5 \pm 0.5)^\circ$ with the Bayesian method and $\gamma_0 = (27.8 \pm 0.8)^\circ$ with the barycenters method.

\section{Discussion} \label{s:discussion}

We model the LOS magnetograms of 21 bipolar emerging ARs observed during Solar Cycle 23. The half-torus model used here provides an estimation of the FR magnetic parameters such as the tilt angle and the twist. We find that the Bayesian method estimates the tilt by correcting the effect of the magnetic tongues. This estimation is therefore a better proxy of the intrinsic inclination of the FR than the estimation obtained using the magnetic barycenters. A proper estimation of the tilt at the early stages of the AR evolution is necessary for a correct analysis of the origin of the tilt dispersion.

The estimation of Joy's law using the corrected tilt angle obtained with the Bayesian method shows differences between the early and mid stages and the last part of the emergence. These differences indicate that the stage of AR emergence has to be considered when Joy's law is derived using statistical analyses. The lower slope and reduced dispersion of Joy's law at the last stage of the emergence suggest that the tilt relaxes to a stable value. The parameters of the fit of the latitudinal dependence of the tilt are fairly in agreement with the ones obtained by \cite{Stenflo2012}, indicating that our sample is within the expected range of observed cases.

Moreover, when applying the Bayesian method we only find six estimations of the tilt that do not follow the expected Joy's orientation against nine cases when using the barycenters. The Bayesian method reduces the amount of anti-Joy's measurements compared to the estimation using the magnetic barycenters. The proper estimation of the tilt in these kinds of regions provides a better tool for a deeper analysis of the origin of these FRs.

\end{document}